\documentclass[reprint,aps,prx]{revtex4-2}
\usepackage[utf8]{inputenc}
\usepackage[T1]{fontenc}
\usepackage{caption}
\usepackage{subcaption}
\usepackage{amssymb,amsmath}
\usepackage{graphicx}
\usepackage{xcolor,tikz}
\usepackage[pdfusetitle]{hyperref}
\usepackage[justification=raggedright]{caption}

\DeclareMathOperator\dist{dist}

\usetikzlibrary{calc, positioning, arrows.meta, decorations.pathreplacing}
\definecolor{solyellow}{HTML}{b58900}
\definecolor{solorange}{HTML}{cb4b16}
\definecolor{solred}{HTML}{dc322f}
\definecolor{solmagenta}{HTML}{d33682}
\definecolor{solviolet}{HTML}{6c71c4}
\definecolor{solblue}{HTML}{268bd2}
\definecolor{solcyan}{HTML}{2aa198}
\definecolor{solgreen}{HTML}{859900}

\definecolor{vir1}{HTML}{440154} 
\definecolor{vir2}{HTML}{472c7a} 
\definecolor{vir3}{HTML}{3b518b} 
\definecolor{vir4}{HTML}{2c718e} 
\definecolor{vir5}{HTML}{21908d} 
\definecolor{vir6}{HTML}{27ad81} 
\definecolor{vir7}{HTML}{5cc863} 
\definecolor{vir8}{HTML}{aadc32} 
\definecolor{vir9}{HTML}{fde725} 

\begin{document}

\title{LightSABRE: A Lightweight and Enhanced SABRE Algorithm}

\author{Henry Zou}
\affiliation{IBM Quantum, Yorktown Heights, NY 10598 USA}
\email{henry.zou@ibm.com}
\author{Matthew Treinish}
\affiliation{IBM Quantum, Yorktown Heights, NY 10598 USA}
\author{Kevin Hartman}
\affiliation{IBM Quantum, IBM Research - Cambridge, Cambridge, MA, USA}
\author{Alexander Ivrii}
\affiliation{IBM Quantum, IBM Research Israel, Haifa, Israel}
\author{Jake Lishman}
\affiliation{IBM Quantum, IBM Research Europe, Hursley, UK}

\hypersetup{
    pdfauthor={Matthew Treinish, Jake Lishman, Kevin Hartman, Alexander Ivrii and Henry Zou},
}

\begin{abstract}
We introduce LightSABRE, a significant enhancement of the SABRE algorithm
that advances both runtime efficiency and circuit quality. LightSABRE addresses the increasing 
demands of modern quantum hardware, which can now accommodate complex scenarios, and circuits with millions of gates. 
Through iterative development within Qiskit, primarily using the Rust programming
language, we have achieved a version of the algorithm in Qiskit
1.2.0 that is approximately 200 times faster than the implementation in Qiskit 0.20.1, which
already introduced key improvements like the release valve mechanism.
Additionally, when compared to the SABRE algorithm presented in Li et al.,
LightSABRE delivers an average decrease of 18.9\% in SWAP gate count across the same benchmark circuits.
Unlike SABRE, which struggles with scalability and convergence on large circuits, LightSABRE 
delivers consistently high-quality routing solutions, enabling the efficient execution of large quantum 
circuits on near-term and future quantum devices. LightSABRE's improvements in speed, scalability, 
and quality position it as a critical tool for optimizing quantum circuits in the context of 
evolving quantum hardware and error correction techniques.

\end{abstract}

\date{\today}

\maketitle

\section{Introduction}\label{sec:intro}

A key step in quantum compilation involves mapping and routing a quantum circuit onto a physical 
device while adhering to the device's connectivity constraints. 
In the process, one aims for circuits \emph{of higher quality}: having a smaller size (the number of gates) 
and/or a smaller depth (the number of layers into which the gates can be partitioned).

The SABRE algorithm~\cite{li:2018} has been widely regarded as a leading solution for this task. 
Published in 2018, it has provided high-quality output circuits with reasonable runtime performance, 
establishing itself as the state of the art for the quantum hardware and circuit sizes available at that time.
However, as quantum computing has advanced, resulting in larger device sizes and more complex circuits, 
the runtime of the original SABRE algorithm has become a significant concern.
For instance, there is a growing need to support control flow, which the original SABRE does not address. 

This paper details the modifications made to the SABRE algorithm within 
Qiskit~\cite{qiskit2024} to address runtime, quality, and other concerns
relevant to an industrial-strength quantum compiler. 
In what follows, we introduce LightSABRE as the enhanced version of SABRE
within Qiskit.

Since SABRE's publication, substantial effort from the community has focused on improving 
the quality of output circuits through various extensions and modifications. 
To cite a few papers (this list is by no means complete):
\cite{ForeSight} keeps track of multiple candidates circuits during routing and continuously adapts this set by replacing 
worse candidates by better ones;
\cite{Huang2024} considers distance-two bridge gates in addition to swap gates;
\cite{QMDLA} presents a look-ahead heuristic that improves the circuit size;
\cite{Cheng_2024} describes a scheme to reduce the circuit depth/execution time;
\cite{DBLP:conf/qce/LiuYWHKI23} combines routing and synthesis.
While many of these works focus primarily on circuit quality improvements, 
they often come with a substantial runtime cost, making them impractical for large-scale circuits
In contrast, LightSABRE's primary advantage lies in its dramatic runtime improvement, 
making it highly scalable for circuits of unprecedented size, while consistently 
delivering significant improvements in circuit quality.

In order to improve performance, LightSABRE is implemented using the Rust programming 
language~\cite{10.1145/2663171.2663188}.
A key mechanism in LightSABRE for achieving high-quality circuits is the 
\emph{relative scoring mechanism}, which evaluates the 
heuristic benefit of selecting a particular candidate swap in $\mathcal O(1)$ time, 
as detailed in Section~\ref{subsec:relative-scoring}. 
This efficient scoring method is crucial for improving performance, 
especially as the algorithm scales to larger circuits.

LightSABRE further enhances circuit quality by exploiting SABRE's nondeterministic 
behavior. In practice  SABRE's output greatly varies depending on the initial random layout 
and the stochastic decisions made during the routing process. LightSABRE runs multiple
trials of the algorithm and selects the highest-quality
output circuits, whether in terms of swap count or depth, see Section~\ref{subseq:multiple-trials}
for details. This makes the runtime of the algorithm particularly crucial. It directly impacts the 
feasibility of running multiple trials of the algorithm and achieving high-quality results.
To improve initial layouts, LightSABRE introduces the ability to seed the layout with 
additional strategies beyond random mapping, potentially providing better starting configurations, 
discussed in Section~\ref{subseq:initial-layout}.

As quantum hardware matures, a modern quantum compiler is required to handle circuits and 
devices with additional features. 
First, there is a need to support architectures whose connectivity graphs consist of multiple 
disjoint components. 
Second, there is a need to support circuits with classical feedforward and control flow. 
Both of these pose a unique set of challenges which we discuss in~\ref{subseq:disjoint-connectivity} 
and~\ref{subseq:control-flow} respectively.

Improving the output circuit quality remains an ongoing goal.
In Section~\ref{subsec:heuristic-enhancements} we introduce two new heuristic enhancements
to the core algorithm based on optimizing the circuit depth and ensuring efficient routing
through critical paths. LightSABRE includes a \emph{release valve} mechanism to resolve 
cases where SABRE's original \texttt{lookahead} heuristic could become stuck, 
ensuring forward progress, as explained in Section~\ref{subsec:release-valve}.

These improvements make LightSABRE capable of efficiently handling large quantum circuits, 
positioning it as a forward-looking solution for the demands of emerging quantum hardware.

\section{Improvements}\label{sec:improvements}

\subsubsection{Relative Scoring}
\label{subsec:relative-scoring}

The SABRE heuristic, as described in ref.~\cite{li:2018}, is:
\begin{equation}\label{eq:lookahead-heuristic}
H =
    \underbrace{\frac1{\lvert F\rvert}\sum_{(i,j)\in F} \dist(i, j)}_{\text{\texttt{basic} component}}
    + \underbrace{\frac{k}{\lvert E\rvert}\sum_{(i,j)\in E} \dist(i, j)}_{\text{\texttt{lookahead} component}},
\end{equation}
where $F$ and $E$ are the sets of gates in the front layer and extended set, 
respectively, and $k$ is a relative weighting chosen by the implementer. 
The two sums are over the pairs of \emph{physical} qubits whose virtual qubits partake 
in a gate in the relevant set. 
The function $\dist(i, j)$ counts the distance between physical qubits $i$ and $j$; 
two qubits that can directly interact have a distance of unity.
The heuristic is a scoring for the total system of the front layer and the extended set under 
the assumption that a lookahead table for $\dist$---which requires only the hardware topology 
to be known---is precalculated.
Calculation of $H$ has a computational complexity of $\Theta(\lvert F\rvert + \lvert E\rvert)$.

If no gates from the front layer are routable, SABRE iterates over candidate swaps, evaluating 
the heuristic in eq.~\eqref{eq:lookahead-heuristic} for the system after the candidate swap 
has been made, and selects randomly from the set of swaps that induce the lowest heuristic value. 

A swap is a candidate if it involves at least one qubit that is an operand of a front-layer gate.
Most hardware topology families have some finite average connectivity for each qubit, no matter 
their dimension: rings have nodes of degree 2; periodic grids have nodes of degree 4; 
the heavy-hex lattice has nodes of degrees up to 3. The number of candidate swaps is, therefore, 
typically proportional to $\lvert F\rvert$.
Assuming the maximum size of the extended set is at most some constant proportion to the maximum 
front-layer size, the computational complexity of choosing a ``best'' swap by this method 
is $\Theta\bigl({\lvert F\rvert}^2\bigr)$.

However, applying a single swap affects a maximum of $2 + \lvert E\rvert$ terms in 
eq.~\eqref{eq:lookahead-heuristic}.
Of the \texttt{basic} component, either one or two gates will have a qubit moved by the candidate swap.
It is possible for the entire extended set to be affected by a single swap, if for example, 
the extended set is filled by a star interaction graph, but this typically does not apply for 
every candidate swap.
The extended set, though, is typically constrained to be either a qubit-count-independent size, 
or to be limited to gates up to some constant maximum two-qubit depth beyond the front layer.
In both these cases---which include Qiskit's implementation---the number of terms in 
eq.~\eqref{eq:lookahead-heuristic} affected by a candidate swap is $\mathcal O(1)$.

The best-swap condition is an argument minimization problem over $H_{i\leftrightarrow j}$, 
where $(i,j)$ is a candidate swap and $H_{i\leftrightarrow j}$ is the heuristic of 
eq.~\eqref{eq:lookahead-heuristic} after the swap has been made.
$H_{i\leftrightarrow j}$ is minimised by the same swap that minimises $H_{i\leftrightarrow j} - H_0$ 
for any constant $H_0$.
The core algorithmic improvement of LightSABRE is to minimise this offset cost function, 
with $H_0$ chosen as the value of the heuristic \emph{without making any swaps}.
This can determined without calculating $H_0$ directly; each swap is scored only by the 
relative change it effects, which requires evaluating only $\mathcal O(1)$ terms.
The computational complexity of choosing the best swap in LightSABRE is thus $\Theta(\lvert F\rvert)$, 
where the scaling comes only from the number of candidate swaps.

\subsubsection{Multiple Trials}
\label{subseq:multiple-trials}

In the original SABRE algorithm, there are two parts of the algorithm 
where stochastic elements are introduced: the initial layout is randomly selected, 
and during routing, when selecting a candidate swap to use, if there are multiple 
candidates with identical minimum scores, the swap is chosen at random. These stochastic 
elements result in the original SABRE algorithm's output quality being potentially very
dependent on the random number generator. To ameliorate this impact the
LightSABRE algorithm introduces multiple trials where the algorithm is run
multiple times in parallel using different random number generator seeds for
each trial. Then among all the trials the output which results in the fewest
number of swap gates is selected as the final result. 

Running the multiple trials in parallel minimizes the impact on runtime. Of particular importance when
running multiple trials is to note that the typical objective function used in
routing of minimizing the swap count does not necessarily result in a better
layout being selected when using routing for layout purposes. For this reason
LightSABRE only runs a single routing trial when running routing as part of
layout, but will run multiple routing trials after a layout has been finalized.


\begin{figure}
  
  \begin{subfigure}{\columnwidth}
      \includegraphics[width=\columnwidth]{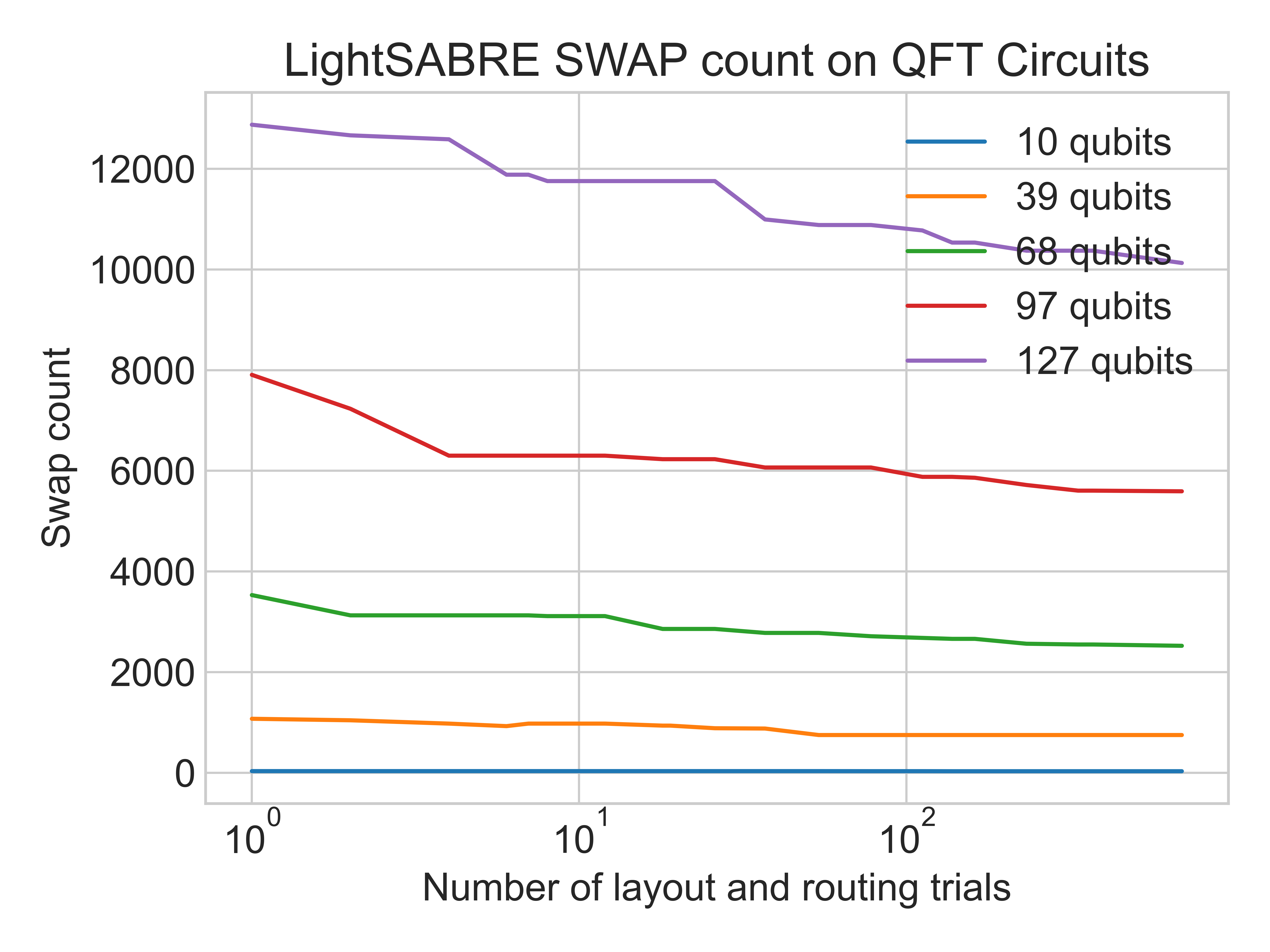}
      \captionsetup{labelformat=empty}
      \caption[FIG. 1a]{FIG. 1a: Swap count for QFT circuits of various sizes targeting a 127 qubit heavy-hex
      backend over multiple LightSABRE trials.
      The swap count consistently decreases or remains constant as the number of trials increases, 
      as LightSABRE selects the trial with the minimum number of swap gates.}

      \label{fig:seed_swap_count}
  \end{subfigure}

  \begin{subfigure}{\columnwidth}
      \includegraphics[width=\columnwidth]{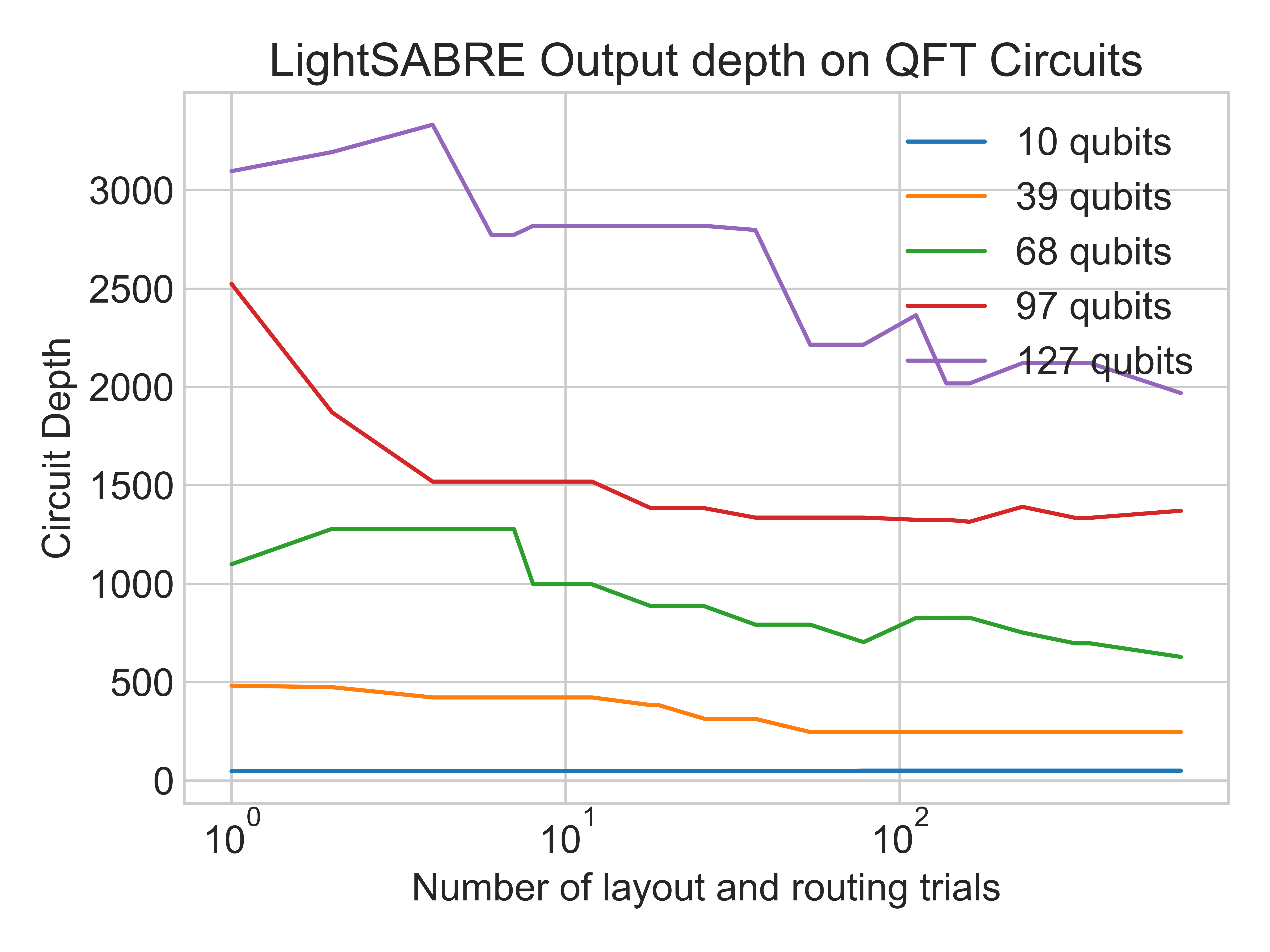}
      \captionsetup{labelformat=empty}
      \caption[FIG. 1b]{FIG. 1b: Circuit depth for QFT circuits of various sizes targeting a 127 qubit heavy-hex
      backend over multiple LightSABRE trials. Circuit depth does not always follow 
      the same trend as swap count, as minimizing swap gates can potentially increase depth. 
      To reduce depth, the objective function could be modified to prioritize depth instead of swap count.}
      \label{fig:seed_depth}
  \end{subfigure}
\end{figure}

Figure \ref{fig:seed_swap_count} demonstrates that increasing the number of layout and routing trials generally leads to better optimization outcomes, particularly in terms of reducing both swap count and circuit depth. 
This trend is especially pronounced for circuits with a larger number of qubits.
As seen in the graphs, for larger QFT circuits, additional trials result in a noticeable reduction in swap count and depth, which are critical factors for optimizing quantum circuits on near-term quantum devices.
Although runtime naturally increases as more trials are added, this increase is moderate and remains manageable even for larger circuits.

\subsubsection{Seeding the Initial Layout}
\label{subseq:initial-layout}

As discussed in the multiple trials section \ref{subseq:multiple-trials}, 
the LightSABRE algorithm performs multiple trials with different initial layouts.
Typically, these starting layouts are selected using a fully random initial mapping, 
similar to the original SABRE algorithm. 
However, LightSABRE also has a provision for manually specifying a list of additional 
starting layouts to use for additional layout trials. 
This option provides the opportunity to improve the initial mapping, 
potentially leading to better overall layout processing than what a fully random 
selection would achieve.
By default, in addition to the fully random trials, 
LightSABRE runs one trial using the most densely connected subgraph of the connectivity graph. 
This additional trial can result in more optimal layouts, particularly for smaller circuits 
being mapped onto larger connectivity graphs.

Qiskit provides several strategies to leverage this mechanism that are available as
analysis passes performed before layout and routing. The VF2Layout analysis pass checks
if the circuit can be perfectly embedded into the device connectivity map by solving the
subgraph isomorphism problem \cite{PRXQuantum.4.010327}.
If this is the case then no further swap mapping or routing is needed. As an example,
a circuit with ring connectivity over 20 qubits can be perfectly mapped into the heavy-hex topology.
The \texttt{SabrePreLayout} analysis pass extends this concept to cases where a perfect layout is not 
possible, yet the circuit can still be mapped ``almost'' perfectly.  
For example, a circuit with ring connectivity over 19 qubits can be mapped onto a 20-qubit ring 
inside a heavy-hex topology, with one qubit missing. 
Similarly, a circuit with ring connectivity over 21 qubits can be mapped 
onto a 20-qubit ring with an additional qubit connected to this ring via a degree-3 vertex. 
In both of these examples the layout is ``almost'' perfect and maps virtually connected
qubits to nearby physical qubits. 

\begin{figure}
  \includegraphics[width=1\columnwidth]{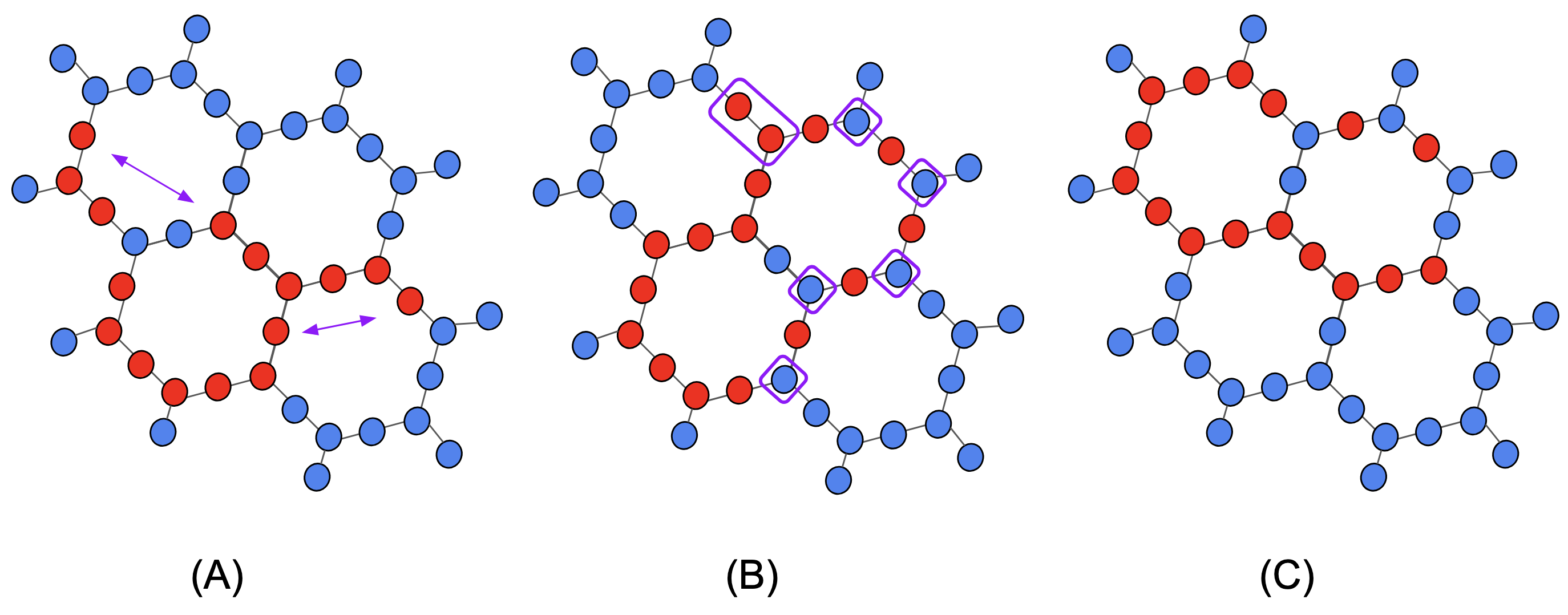}
  \caption{\label{fig:initial-layout}%
  This figure illustrates the effectiveness of the techniques in LightSABRE compared to the 
  original SABRE algorithm, using coupling maps from experiments on the 16-qubit \texttt{EfficientSU2} 
  example with circular entanglement. Panel (A) shows the layout generated by running SABRE, 
  where certain pairs of qubits that are connected in the abstract circuit are separated by 
  significant distances in the physical circuit, as indicated by the purple arrows. 
  Panel (B) shows the layout produced by running \texttt{SabrePreLayout} before SABRE, resulting 
  in a better configuration where all connected nodes are at most distance-2 apart in 
  the physical map. However, this layout remains suboptimal, with a qubit isolated from 
  a pair of red qubits (highlighted with a box) and gaps in the boxed blue qubits. 
  Finally, panel (C) displays the optimal layout achieved by employing the additional 
  minimization feature in \texttt{SabrePreLayout}, which eliminates these inefficiencies and 
  improves overall qubit connectivity.
  }%
\end{figure}

The \texttt{SabrePreLayout} works by augmenting the connectivity graph with additional edges connecting 
pairs of nodes that are within a certain distance $d$ in the original graph 
(typically $d$ is chosen to be $2$). It then solves the subgraph isomorphism problem, using rustworkx~\cite{treinish:2021}, to
determine if a mapping to this augmented connectivity graph exists. Additionally, 
the pass can minimize the number of longer-distance edges by solving further subgraph 
isomorphism problems.

We have observed that as device connectivity maps increase in size, the quality of fully random 
initial layouts tends to deteriorate significantly. Therefore, improving the selection of starting
layouts is an area of active ongoing research \cite{Elsayed_2024}.

\subsubsection{Disjoint Connectivity Graphs}
\label{subseq:disjoint-connectivity}

One underlying assumption of many layout and routing algorithms is that the 
target device's connectivity graph is fully connected, meaning there exists 
a path in the connectivity graph between any two qubits. However, several hardware
vendors have announced \cite{Gambetta2022} modular architectures with multiple
QPUs that have shared classical resources with no quantum connectivity. In such architectures the
connectivity graph is disjoint, and the original SABRE algorithm would not
function correctly, as it assumes there is always a path on
the connectivity graph between all qubits. With disjoint connectivity,
qubits on separate components have no path of swaps that can connect them.
Another scenario when this can occur is if there are any faulty qubits or
2q gates that do not function correctly. When filtering the connectivity
graph to remove potential non-functional qubits this can sometimes result in
a disjoint graph requiring the layout and routing pass to be able to work
with this constraint.

To support this target LightSABRE algorithm intoduces initial analysis and decomposition
to the layout and routing problem. First the connectivity graph is analyzed to find
the connected components. If there is more than one connected component in the graph
then the directed acyclic graph representation of the circuit is similarly also
analyzed for its connected components. A greedy placement algorithm is then used to map 
each circuit connected component onto a connected component in the connectivity graph. 
Following this, the normal layout procedure is run on each connected component 
in isolation, and the layout components are subsequently combined to generate a complete layout for the circuit. 
Routing should be performed after the complete layout is applied; 
running it on an isolated component may potentially yield invalid results, 
as the connected components can only account for the dependency ordering of 
quantum operations. Any dependencies outside of this, such as classical bit 
reuse or control flow, will not be captured when routing is performed on each 
component in isolation.

\subsubsection{Classical Control Flow}
\label{subseq:control-flow}

Qiskit supports circuits with classical control flow, a feature increasingly 
offered at the hardware level by modern devices. 
These circuits expect to perform mid-circuit measurements and select between
one of two or more branches at runtime.
The original SABRE algorithm does not account for control flow in circuits, 
making it unsuitable for routing such circuits. To address this limitation, 
we have extended LightSABRE to include special handling for control flow operations.

\begin{figure}
    \newsavebox{\genericfilt}
\savebox{\genericfilt}{%
    \begin{tikzpicture}[
            qubit/.style={draw, fill=blue!10, circle, minimum size=0.6cm},
            cbit/.style={draw, fill=red!10, circle, minimum size=0.6cm},
            gate/.style={draw, fill=green!20, circle, minimum size=0.5cm},
            edge/.style={->, thick},
            label/.style={midway, fill=white, inner sep=1pt},
            node distance=1cm,on grid
        ]
        \node[draw] (title) at (0,0) {if-else};
        \node[qubit, below right=1cm of title] (q1_1) {$q_1$};
        \node[gate, below=1cm of q1_1] (h) {h};

        \draw [dashed] (0,-0.5) -- (0,-1.5);

        \node[qubit, below left=1cm of title] (q1_2) {$q_1$};
        \node[gate, below=1cm of q1_2] (x) {x};

        \draw[edge] (q1_1.south) -- (h.north) node[label] {$q_1$};
        \draw[edge] (q1_2.south) -- (x.north) node[label] {$q_1$};
    \end{tikzpicture}%
}

\begin{tikzpicture}[
  every node/.style={font=\footnotesize},
  qubit/.style={draw, fill=blue!10, circle, minimum size=0.8cm},
  cbit/.style={draw, fill=red!10, circle, minimum size=0.8cm},
  gate/.style={draw, fill=green!20, circle, minimum size=1.0cm},
  edge/.style={->, thick},
  label/.style={midway, fill=white, inner sep=1pt},
  node distance=1cm,on grid
]


\node[qubit] (q0) at (1,3) {$q_0$};
\node[qubit, right= of q0] (q1) {$q_1$};
\node[cbit] (c0) at (0,3) {$c_0$};
\node[cbit, left= of c0] (c1) {$c_1$};

\node[gate, below=2cm of q0] (h1) {h};
\node[gate, below=4cm of c0, xshift=0cm] (measure1) {meas};

\draw (2,-4) node[draw](ifelse){\usebox{\genericfilt}};

\node[gate, below=3cm of ifelse, xshift=-3cm] (measure2) {meas};

\draw[edge] (q0.south) -- (h1.north) node[label] {$q_0$};
\draw[edge] (h1.south) -- (measure1.north east) node[label] {$q_0$};
\draw[edge] (measure1.south) -- (ifelse.north west) node[label] {$c_0$};
\draw[edge] (q1.south) -- (ifelse.north) node[label] {$q_1$};
\draw[edge] (c0.south) -- (measure1.north) node[label, pos=0.6] {$c_0$};
\draw[edge] (c1.south) -- (measure2.north) node[label, pos=0.6] {$c_1$};
\draw[edge] (ifelse.south) -- (measure2.north east) node[label, pos=0.6] {$q_1$};

\end{tikzpicture}%
    \caption{\label{fig:control-flow}%
    A DAG representation of a circuit which applies an H-gate onto qubit \texttt{q0}, measures the result into
    classical bit \texttt{c0}, and then conditionally applies either an X-gate or H-gate onto qubit \texttt{q1} based
    on the result at runtime. The \texttt{if-else} operation is represented as a gate with data dependencies
    on \texttt{c0} used by its condition and \texttt{q1} used by the inner two circuits of its branches.
    Finally, \texttt{q1} is measured to \texttt{c1} in the outer circuit.
    }%
\end{figure}
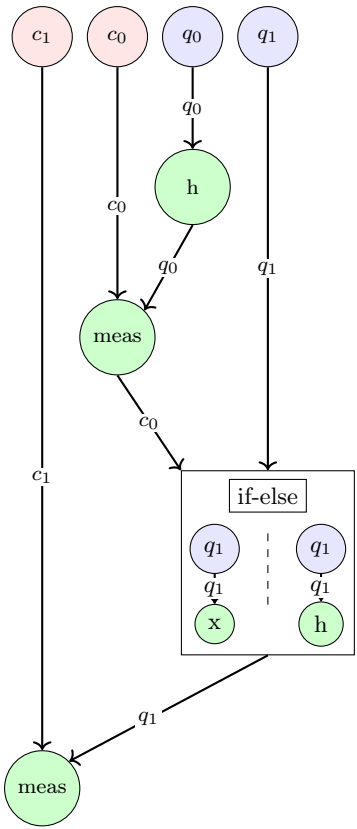

In Qiskit, control flow operations are represented as multi-qubit operations, 
each containing separate circuits for each branch of the associated operation. 
For example, an \texttt{if-else} operation contains two individual circuits:
one for the \texttt{if} block and one for the \texttt{else} block. Similar to how we handle
1Q and 2Q operations, we define the \texttt{if-else} operation to act on the bits it uses,
which is the union of bits used in each of its two circuits
(both quantum and classical) as well as any additional classical bits used by its predicate. 
The \texttt{if-else} operation can only be executed once all predecessor operations
affecting these bits have been executed. Figure~\ref{fig:control-flow} provides a visual
example of these dependencies encoded as a directed acyclic graph (DAG).
This representation is identical to the DAG-based dependency graph used by the original
SABRE algorithm, but encodes control flow operations as multi-qubit gates.

Incorporating control flow operations into the dependency graph has significant 
implications for SABRE\@. In the original SABRE algorithm, the main loop adds gates 
to the front layer as they become executable, i.e.\ once their predecessors have 
been executed. This front layer is then used to evaluate the suitability of candidate 
SWAPs, employing a heuristic that rewards choices aligning the circuit's layout 
with the device's connectivity constraints. However, unlike 2Q gates, 
there is no simple layout choice that can make an arbitrary control flow operation 
executable on a real device. Therefore, these operations do not fit naturally into 
the front layer. Since each branch of a control flow operation is itself a full circuit, 
routing must be performed separately for each branch.

The first consideration we address in LightSABRE is the execution of control flow 
operations as soon as they are encountered by recursively running the algorithm on 
each branch of the operation. This involves adding special handling for control flow 
operations, allowing the routing of blocks recursively as soon as they become executable, 
thereby bypassing the front layer entirely. Each block is routed using an initial layout, 
that matches the outer circuit's layout at the time the control flow operation is encountered. 
While this recursive step is a natural extension of the original algorithm, it can result in 
different final layouts for each branch, which poses a problem for gates following the control 
flow operation. Moreover, it disrupts the `lookahead' mode described by the original algorithm, 
which leverages an extended set of upcoming gates to reward SWAP choices that benefit not only 
the front layer but also subsequent gates.

To address this, our second consideration is ensuring that gates following a control flow 
operation maintain the same layout regardless of which branch is taken at runtime. 
This is achieved by appending an epilogue of SWAPs to each branch to align all branches with a common layout. 
If we choose this common layout to be the same as the starting layout at the time the control flow operation 
was encountered, the \texttt{lookahead} mode remains undisturbed. As a result, LightSABRE's \texttt{lookahead}
will effectively ``look through'' control flow operations as if they were 1Q gates, 
since they do not affect routing from the perspective of the outer circuit.

\subsubsection{Heuristic Enhancements}
\label{subsec:heuristic-enhancements}

In addition to the basic and lookahead components of the heuristic defined in equation 
\eqref{eq:lookahead-heuristic}, we introduce two new heuristic enhancements: \textit{depth} and \textit{critical path}. 
These enhancements provide greater flexibility and optimization in the layout and routing process by enabling the algorithm 
to consider additional factors that impact overall circuit performance. The weights of each heuristic component can be adjusted 
to use either a constant weight or a weight that scales with the size of the set, and each component supports relative scoring, 
offering further customization. These components are entirely independent, allowing for combinations such as depth with critical 
path or just depth alone. Traditionally, SABRE has focused primarily on minimizing the swap count. 
However, with these new components, users can prioritize other factors, such as circuit depth, depending on their optimization goals.

\paragraph{Depth Component}
The depth component introduces a term to the heuristic that aims to reduce the overall circuit depth. 
By incorporating a detailed state tracking mechanism, the algorithm can efficiently update and backtrack, 
considering the depth of each qubit to prioritize paths that minimize the overall increase in circuit depth. 
The dynamic adjustment to the evolving circuit layout ensures that each decision contributes toward the overall optimization goals. 
The depth component is defined as:
\begin{equation} \frac{D \Delta_{depth}}{3} \end{equation}
where $D$ is the weight, and $\Delta_{depth}$ is the difference in depth between the current circuit 
and the circuit after applying the SWAP gate and the immediate routable gates after, with $\Delta_{depth}$.
Note that the depth here represents the 2-qubit gate depth, and the division by 3 accounts for 
each SWAP being represented by three CNOT gates. Although this heuristic can effectively reduce circuit depth, 
it comes with a trade-off in runtime, as the algorithm must track qubit depth after each swap candidate and compute the true depth impact, 
including any subsequent routable gates.

\paragraph{Critical Path Component}
The critical path component introduces a term to the heuristic that prioritizes swaps facilitating the execution 
of critical paths in the circuit. Unlike depth, this component operates on the abstract circuit, 
meaning the critical path does not need to be recomputed after each gate is routed. 
Instead, it tracks the number of descendants of each gate and assigns them a ranking, 
allowing the algorithm to prioritize gates that are critical to the circuit's execution. 
The ranking reflects the gate's depth in the critical path, with rank 1 indicating the gate with the highest depth. 
The critical path component is defined as:
\begin{equation} \alpha^{r_{gate}} \end{equation}
where $r_{gate}$ is the rank of the gate in the critical path, and $\alpha$ is a constant between 0 and 1. 
This term adjusts the overall heuristic, ensuring that gates on the critical path are given higher priority in the optimization process. 
While this heuristic is most effective for circuits with a well-defined critical path, 
it generally produces results similar to the basic heuristic unless the critical path term is given significant weighting.

These heuristic enhancements enable the LightSABRE algorithm to make more informed decisions, 
improving the efficiency and quality of the qubit layout and routing process. 
By incorporating considerations of circuit depth and critical paths, 
LightSABRE is better equipped to handle the increasing complexity and size of modern quantum circuits.

\paragraph{Comparison of Heuristics}
The performance of the various heuristics integrated into LightSABRE demonstrates that no single heuristic universally outperforms the others 
across all metrics or circuit types. The choice of heuristic should be tailored to the specific optimization goals and 
the nature of the quantum circuit being optimized. It is challenging to strictly compare the heuristics as 
they perform differently depending on the circuit structure.

\begin{figure}
  
  \begin{subfigure}{\columnwidth}
      \includegraphics[width=1\columnwidth]{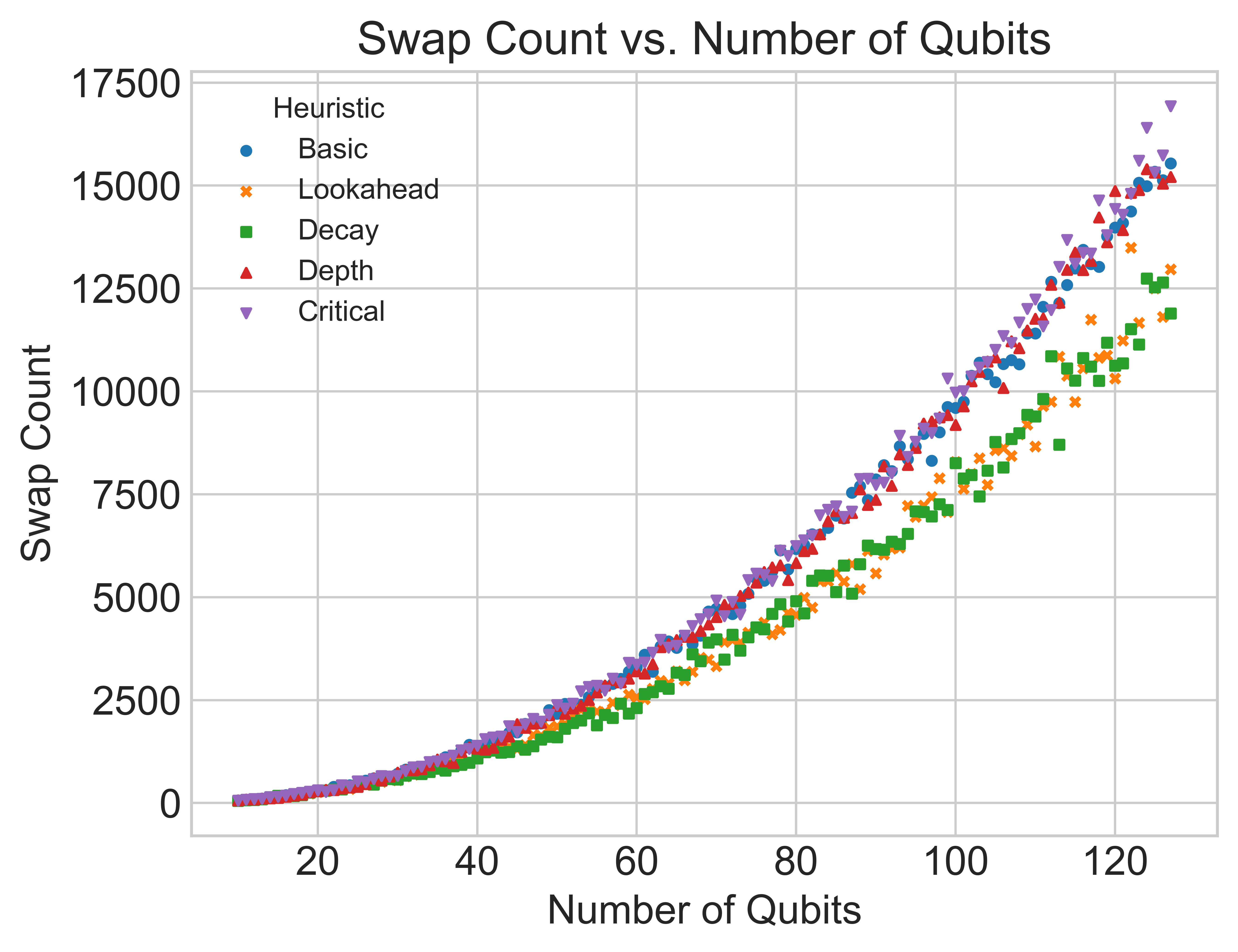}
      \captionsetup{labelformat=empty}
      \caption[FIG. 4a]{FIG. 4a: Swap count for QFT circuits of various sizes targeting a 127 qubit heavy-hex
      backend, comparing multiple heuristics. 
      The lookahead and decay heuristics perform the best in terms of reducing swap count, 
      with approximately 15\% fewer swaps than the basic heuristic.}
      \label{fig:heuristics_qft_swap}
  \end{subfigure}

  \begin{subfigure}{\columnwidth}
      \includegraphics[width=1\columnwidth]{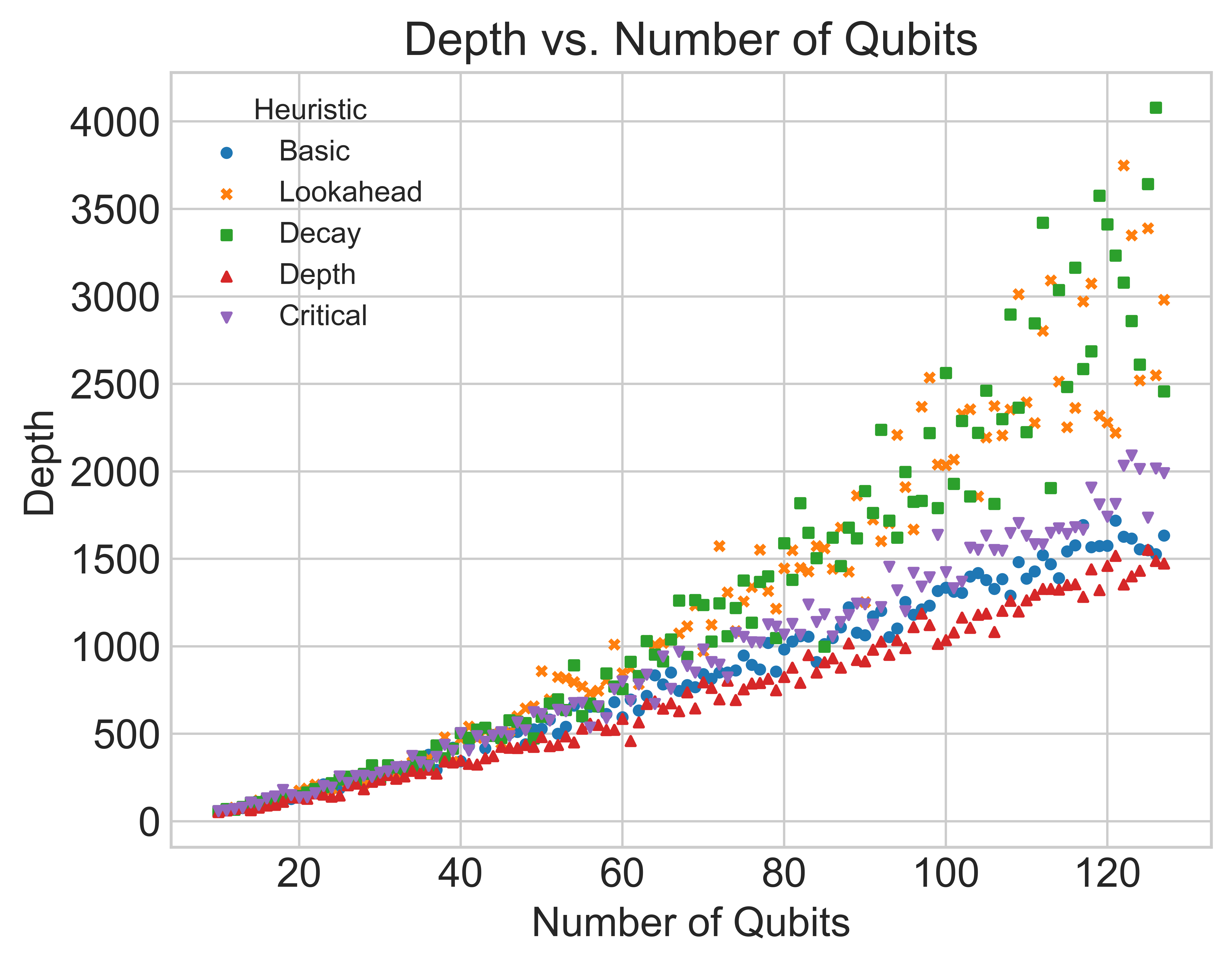}
      \captionsetup{labelformat=empty}
      \caption[FIG. 4b]{FIG. 4b: Circuit depth for QFT circuits of various sizes targeting a 127 qubit heavy-hex
      backend, comparing multiple heuristics. 
      While lookahead and decay heuristics reduce swap count, they consistently 
      lead to nearly twice the circuit depth compared to the basic heuristic. 
      The depth heuristic performs best by maintaining lower depths across trials.}
      \label{fig:heuristics_qft_depth}
  \end{subfigure}
\end{figure}

In Figures \ref{fig:heuristics_qft_swap} and \ref{fig:heuristics_qft_depth}, 
we observe the trade-offs between different heuristics for the QFT circuit. 
From the swap count graph (Fig. \ref{fig:heuristics_qft_swap}), 
we see that the \texttt{lookahead} and \texttt{decay} heuristics are the most 
effective for reducing swap count. However, when examining the circuit depth in
 Fig. \ref{fig:heuristics_qft_depth}, we see that these same heuristics result 
 in nearly double the depth compared to the \texttt{basic} heuristic, 
 highlighting their inefficiency in managing circuit depth.
In contrast, the \texttt{depth} heuristic, while not producing the lowest swap count, 
offers a balanced trade-off by maintaining a similar swap count to the \texttt{basic} 
heuristic while consistently achieving lower circuit depths. This suggests that 
the \texttt{depth} heuristic could be considered the best overall heuristic for 
this specific scenario, as it provides reasonable performance in both metrics.

It is important to note that this analysis is based on the QFT circuit, 
and performance may vary for other types of circuits. However, given the underlying 
characteristics of the heuristics, it is likely that similar trends will be observed 
for other circuits. Users can always test and identify the most suitable heuristic 
for their specific circuits and optimization goals. 

One trade-off to consider with the \texttt{depth} heuristic is the increased runtime, 
as it requires tracking qubit depth and computing the true impact of each swap 
candidate on the circuit depth.

To gain a better understanding of how these heuristics perform across a wide range of circuits, 
we tested them on 500 different Quantum Volume circuits, ranging from 10 qubits to 50, 
with original depths ranging from 10 to 25. While it is challenging to construct a circuit benchmark 
that encompasses a vast range of different circuit types, Quantum Volume circuits serve as an excellent starting point. 
These circuits are particularly difficult to route for any routing pass, and their random nature allows them to cover a 
broad spectrum of circuit configurations. This makes them an ideal choice for evaluating the robustness 
and effectiveness of our heuristic enhancements.

\begin{figure}
  
  \begin{subfigure}{\columnwidth}
      \includegraphics[width=\columnwidth]{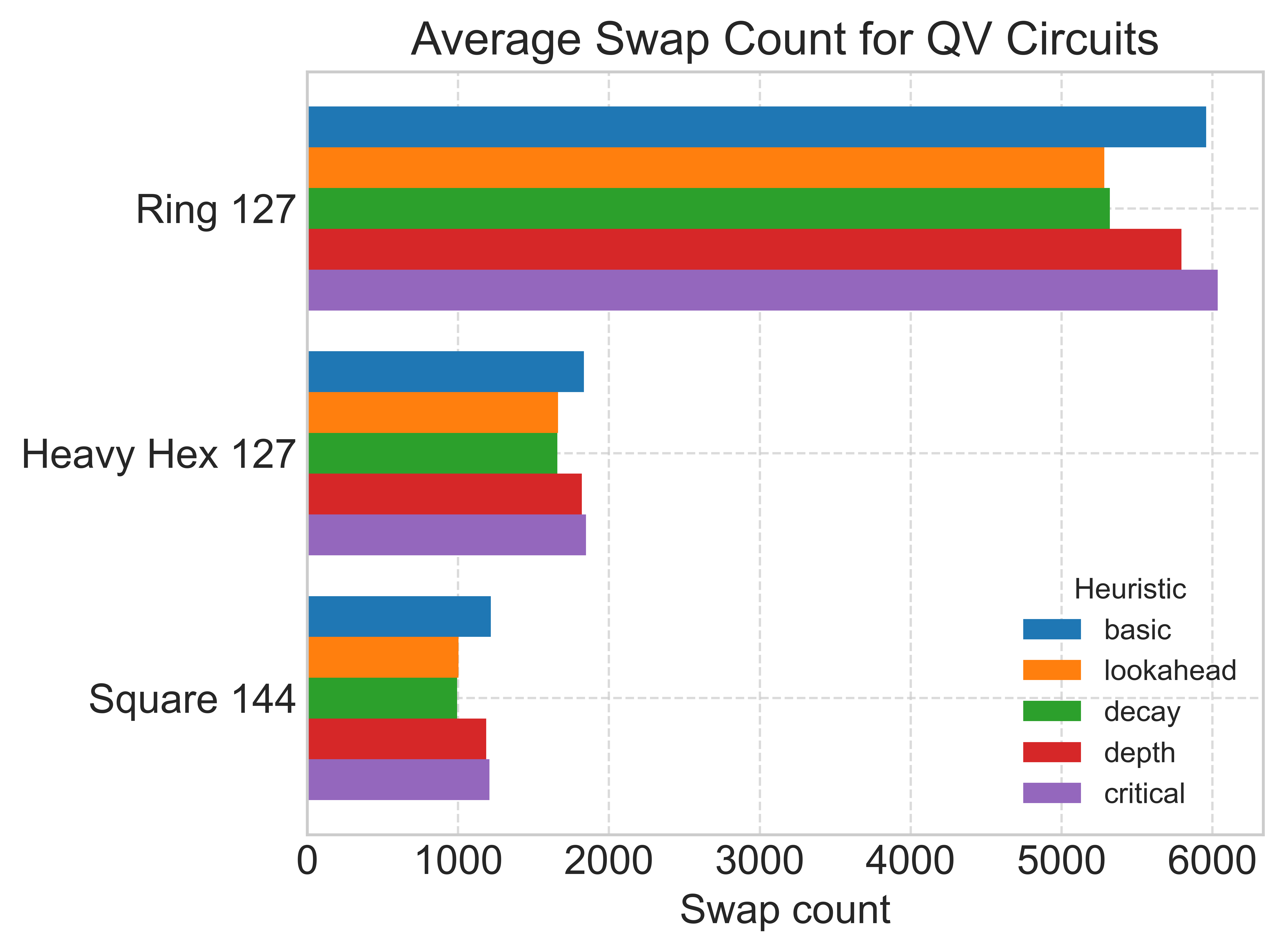}
      \captionsetup{labelformat=empty}
      \caption[FIG. 5a]{FIG. 5a: Average swap counts for QV circuits across three coupling maps. 
      Each heuristic was ran with the same seed list and 20 swap trials for each run. 
      The heuristics provide minimal difference in swap count, 
      likely due to the complexity of routing QV circuits.}
      \label{fig:cmap_swap}
  \end{subfigure}
  \begin{subfigure}{\columnwidth}
    \includegraphics[width=\columnwidth]{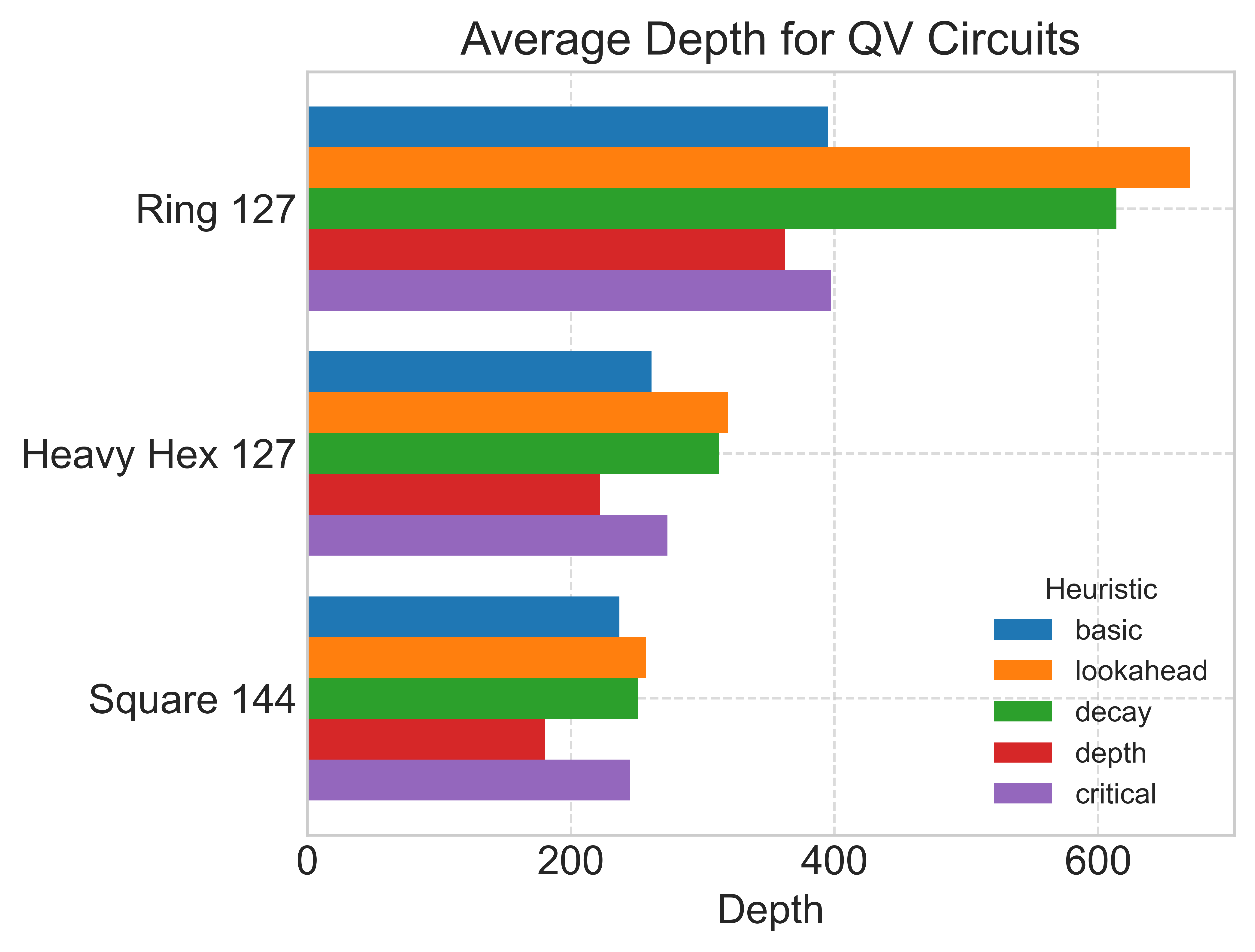}
    \captionsetup{labelformat=empty}
    \caption[FIG. 5b]{FIG. 5b: Average circuit depths for Quantum Volume circuits across three coupling maps. 
    Each heuristic was ran with the same seed list and 20 swap trials for each run. 
    Lookahead and decay heuristics significantly increase depth compared to the basic heuristic, 
    but the impact is less pronounced with higher-connectivity coupling maps like heavy xex and square.}
    \label{fig:cmap_depth}
\end{subfigure}

\end{figure}

Experiments across various circuit QVOLs reveal that while lookahead and decay generally offer the best reduction in swap counts, 
the depth heuristic usually provides substantial depth reductions, whereas the critical path heuristic is 
more effective in the few cases where the critical path is clearly evident.
These findings are consistent across different coupling maps, though the differences between heuristics 
diminish as the connectivity of the map increases. 
This suggests that as device connectivity improves, the choice of heuristic may become less critical for achieving optimal results, 
though it remains relevant for tailored circuit optimizations.

Overall, these findings highlight the importance of selecting the appropriate heuristic based on the 
specific goals of the quantum circuit optimization. 
By adding these new components, LightSABRE provides users with the ability to tailor 
the optimization process to their needs, 
whether that be minimizing swap count, reducing depth, lowering runtime, or prioritizing critical path execution.
While each heuristic has its strengths and weaknesses, their inclusion in LightSABRE allows for a more versatile and 
adaptive approach to quantum circuit optimization. 
This flexibility is crucial as quantum circuits continue to grow in complexity, and different 
applications demand different optimization criteria. 
By testing and selecting the best heuristic for a given metric, users can ensure that 
their circuits are optimized according to the most relevant parameters for their specific use case.

\subsubsection{Release Valve}
\label{subsec:release-valve}

\begin{figure}
  \begin{tikzpicture}[%
  every node/.style={
    inner sep=0mm,
  },
  track/.style={
    draw=black,
  },
  layer box/.style={
    draw=#1,
    dashed,
    rounded corners,
  },
  gate/.style={
    draw=#1,
    thick,
    shorten <=-2.5pt,
    shorten >=-2.5pt,
    {Circle[length=5pt]}-{Circle[length=5pt]},
  },
  cost brace/.style={
    draw,
    decorate,
    decoration={brace,raise=4pt, amplitude=4pt},
  },
  cost node/.style={
    midway,
    right=10pt,
  },
]
\makeatletter
\newlength\inf@qubitsep\setlength\inf@qubitsep{5mm}
\newlength\inf@layersep\setlength\inf@layersep{8mm}
\newcommand*\inf@maxlayer{5}
\newcommand*\inf@maxqubit{12}

\foreach \q in {0, 1, ..., \inf@maxqubit} {
  \coordinate (q\q-l) at ($ (0, -\q * \inf@qubitsep) $);
  \foreach \i in {0, 1, ..., \inf@maxlayer} {
    \coordinate (q\q-layer\i) at ($ (q\q-l) + 0.5*(\inf@layersep, 0) + \i *(\inf@layersep, 0) $);
  }
  \coordinate (q\q-r) at ($ (q\q-layer\inf@maxlayer) + 0.5*(\inf@layersep, 0) $);

  \path [track] (q\q-l) node [left=2mm] {$q_{\q}$} -- (q\q-r);
}
\path [gate=black] (q0-layer0) -- (q12-layer0);
\path [gate=black] (q1-layer1) -- (q11-layer1);
\path [gate=black] (q2-layer2) -- (q10-layer2);
\path [gate=black] (q3-layer3) -- (q9-layer3);
\path [gate=solblue] (q5-layer4) -- (q7-layer4);
\path [gate=black] (q4-layer5) -- (q5-layer5);
\path [gate=black] (q7-layer5) -- (q8-layer5);

\newcommand*\inf@layerbox[4]{%
  \path let \p1 = (0.4\inf@layersep, 0.4\inf@qubitsep) in 
    [layer box=#1] ([shift={(-\x1,\y1)}]#2) rectangle ([shift={(\x1,-\y1)}]#3);
  \path (#3 -| #2) -- (#3) node [midway, below=0.6\inf@qubitsep, color=#1] {#4};
}
\inf@layerbox{solblue}{q0-layer0}{q12-layer4}{Front layer};
\inf@layerbox{solred}{q0-layer5}{q12-layer5}{Extended set};

\path [cost brace] (q0-r) -- (q3-r) node [cost node] {$0$};
\path              (q3-r) -- (q4-r) node [cost node] {$w - 1$};
\path              (q4-r) -- (q5-r) node [cost node] {$1$};
\path [cost brace] (q5-r) -- (q7-r) node [cost node, name=midlabel] {$w - 1$};
\path              (q7-r) -- (q8-r) node [cost node] {$1$};
\path              (q8-r) -- (q9-r) node [cost node] {$w - 1$};
\path [cost brace] (q9-r) -- (q12-r) node [cost node] {$0$};
\node [right=3mm of midlabel, anchor=north, rotate=90] {Cost of making swap};

\makeatother
\end{tikzpicture}%
%
  \caption{\label{fig:infinite-circuit}%
    Simplified interaction graph laid out on a linear topology on which the \texttt{lookahead} heuristic is unable to make forwards progress.
    Gates requiring routing are marked by joined circles.
    Swap costs are indicated on the right edge; braces denote that all neighboring qubits within the brace have the same swap cost.
    The blue gate needs only one swap to be routed, but its cost is $w-1$, where $w = k\lvert F\rvert/\lvert E\rvert$ is the relative weighting of the unit distance within the extended set as compared to within the front layer with respect to eq.~\eqref{eq:lookahead-heuristic}.
    If the front layer has many gates compared to the extended set, the necessary swaps can be more costly than shuffling the outermost qubits without making progress.
  }%
\end{figure}
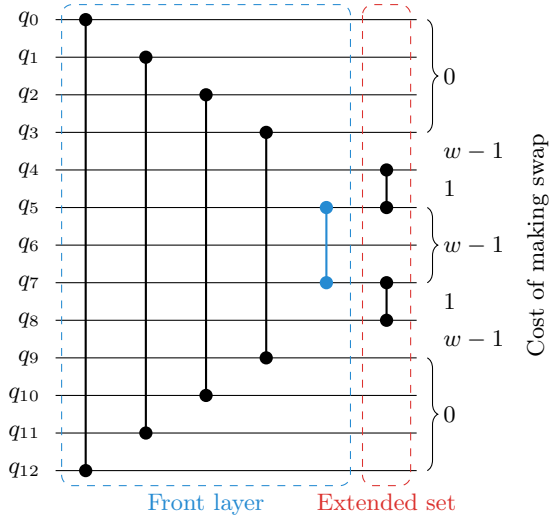

The Sabre heuristic of eq.~\eqref{eq:lookahead-heuristic} is susceptible to getting stuck in a local minimum, where the best-scoring swaps will never make forward progress.
Figure~\ref{fig:infinite-circuit} illustrates a simple case where this heuristic can fail to make forwards progress in any amount of time, if the only swaps considered are those with the best score.
The physical qubits are laid out in a linear nearest-neighbor topology, and blue and red boxes mark the front layer and extended set, respectively.
The heuristic difference in making a swap that brings a single gate in the front layer one step closer together is $-1/\lvert F\rvert$, while the cost of pulling apart a gate in the extended set is $k/\lvert E\rvert$.
Consequently, if $k > \lvert E\rvert/\lvert F\rvert$, which Qiskit's choice of $k=\frac12$ satisfies for the circuit of fig.~\ref{fig:infinite-circuit}, the best swaps will always be those that swap any pair of ``outer'' qubits.
These make no progress towards routing the blue gate, whose qubits are prevented from moving by a heuristic hill imposed by the gates of the extended set.

A circuit with a similar structure to fig.~\ref{fig:infinite-circuit} can be made arbitrarily hard for the \texttt{lookahead} heuristic.
The more ``outer'' qubits that have long-range gates are in the front layer, the lower the threshold value for $k$ is at which the algorithm can fail to make progress.
If the swap-selection routine is extended to allow uphill swaps with lesser probability, the overall chance of finding the path to route the gate can be lowered exponentially by increasing the number of unused qubits between $q_5$ and $q_7$.
Circuits with both of these additional complications have been found in real-world applications of Qiskit's transpiler targetting hardware with topologies such as heavy-hex.

LightSABRE keeps track of the swaps it has made since the last time a gate was routed.
If the number of these exceeds some heuristically chosen threshold, the algorithm considers itself stuck, and backtracks by reapplying the swaps to its current state in reverse order.
This returns the state to the point at which the last gate was routed.
From here, the algorithm uses Dijkstra's algorithm~\cite{Dijkstra1959} to find the shortest path between between the current physical qubits of the gate in the front layer that has the smallest distance between its operands.
Swaps are applied from both ends of the path to cause the qubits to meet in the middle.
This gate is then routed, and the LightSABRE algorithm proceeds as normal.

Backtracking and greedily routing a single gate is not an efficient routing strategy in the general case.
We find that this situation is only very infrequently necessary in real-world circuits, however, and so our implementation is focussed on having zero runtime cost when not needed, and providing a completely fail-safe ``release valve'' mechanism to escape from an arbitrarily deep heuristic local minimum.

\section{Benchmarking results}\label{sec:benchmarking}

To evaluate the performance of the LightSABRE algorithm, we conducted a series of benchmarking experiments, 
comparing its results against those produced by the original SABRE algorithm as presented in Li et al.~\cite{li:2018}. 
The primary metric used in these benchmarks is the CNOT gate count, which directly impacts the fidelity and execution time of quantum circuits on hardware.

Table \ref{tab:benchmark} presents the results from running the same set of benchmark circuits used in Li’s paper. 
The results demonstrate a significant reduction in the number of CNOT gates, 
with an average decrease of 18.9\% across the benchmark circuits.
\begin{table}[h!]
\resizebox{0.49\textwidth}{!}{ 
  \begin{tabular}{|l|l|l|l|l|l|}
    \hline
    Circuit Name & $g_{O}$ & $\bar{g_{LO}} \pm \sigma(g_{LO})$ & $\bar{g_{LD}} \pm \sigma(g_{LD}) $ & $\frac{g_O - \bar{g_{LD}} }{g_O}$ & $\frac{\bar{g_{LO}}  - \bar{g_{LD}} }{\bar{g_{LO}} }$ \\
    \hline
    qft\_10 & 54 & 37.4 $\pm$ 4.6 & 32.3 $\pm$ 2.4 & -40.1\% & -13.6\% \\
    qft\_16 & 186 & 158.9 $\pm$ 10.2 & 134.3 $\pm$ 6.5 & -27.8\% & -15.5\% \\
    rd84\_142 & 105 & 122.0 $\pm$ 13.5 & 99.3 $\pm$ 9.6 & -5.4\% & -18.6\% \\
    adr4\_197 & 1614 & 1375.0 $\pm$ 113.6 & 1092.7 $\pm$ 56.0 & -32.3\% & -20.5\% \\
    radd\_250 & 1275 & 1311.2 $\pm$ 102.4 & 1071.8 $\pm$ 61.5 & -15.9\% & -18.3\% \\
    z4\_268 & 1365 & 1239.5 $\pm$ 96.2 & 954.6 $\pm$ 57.1 & -30.1\% & -23.0\% \\
    sym6\_145 & 1272 & 1407.0 $\pm$ 136.0 & 1088.5 $\pm$ 86.7 & -14.4\% & -22.6\% \\
    misex1\_241 & 1521 & 1566.9 $\pm$ 203.8 & 1037.4 $\pm$ 147.0 & -31.8\% & -33.8\% \\
    rd73\_252 & 2133 & 2314.2 $\pm$ 175.2 & 1913.7 $\pm$ 101.5 & -10.3\% & -17.3\% \\
    cycle10\_2\_110 & 2622 & 2715.5 $\pm$ 155.2 & 2289.4 $\pm$ 99.8 & -12.7\% & -15.7\% \\
    square\_root\_7 & 2598 & 2752.5 $\pm$ 223.5 & 2123.0 $\pm$ 160.8 & -18.3\% & -22.9\% \\
    sqn\_258 & 4344 & 4337.5 $\pm$ 293.1 & 3574.4 $\pm$ 142.0 & -17.7\% & -17.6\% \\
    rd84\_253 & 6147 & 6236.8 $\pm$ 281.2 & 5615.6 $\pm$ 162.5 & -8.6\% & -10.0\% \\
    co14\_215 & 8982 & 8505.7 $\pm$ 372.9 & 7494.6 $\pm$ 339.7 & -16.6\% & -11.9\% \\
    sym9\_193 & 16653 & 16716.7 $\pm$ 643.6 & 15214.0 $\pm$ 886.6 & -8.6\% & -9.0\% \\
    9symml\_195 & 17268 & 16716.7 $\pm$ 643.6 & 15214.0 $\pm$ 886.6 & -11.9\% & -9.0\% \\
    \hline
    \hline 
    Average & &  &  & \textbf{-18.9\%} & \textbf{-17.4\%} \\
    \hline
    \end{tabular}
}
\caption{Benchmark comparison of original SABRE, LightSABRE with SABRE's configuration, 
and LightSABRE's default. $g_{O}$ represents the CNOT gates added by original SABRE, 
$\bar{g_{LO}}$ is the average added by LightSABRE using SABRE's configuration, 
and $\bar{g_{LD}}$ is the average with LightSABRE's default settings. 
LightSABRE results are averaged over 50 runs using Qiskit 1.2.0, while original SABRE uses a single run. 
For $LO$, LightSABRE was set to \texttt{swap\_trials=1, layout\_trials=5, max\_iterations=3, heuristic=decay} 
to match SABRE. For $LD$, default settings were used: 
\texttt{swap\_trials=20, layout\_trials=20, max\_iterations=4, heuristic=decay}. 
LightSABRE with SABRE's settings generally aligns with original SABRE, 
though minor deviations in $g_{O}$ may occur, 
which may suggest slight variances due to other changes in LightSABRE.}
\label{tab:benchmark}
  \end{table}

The primary enhancement of LightSABRE is its improved runtime efficiency, 
but the benchmark results also demonstrate significant improvements in circuit quality. 
This is evident from the reduced CNOT gate count compared to both the original 
SABRE and LightSABRE using the original SABRE configuration. These improvements 
are mainly due to the use of multiple trials in the layout and routing phases, 
allowing LightSABRE to explore more potential solutions and choose the one 
that minimizes gate count most effectively.

\subsection{Scaling Performance}
\label{subsec:scaling-performance}

\begin{figure}[h!]
    \includegraphics[width=\columnwidth]{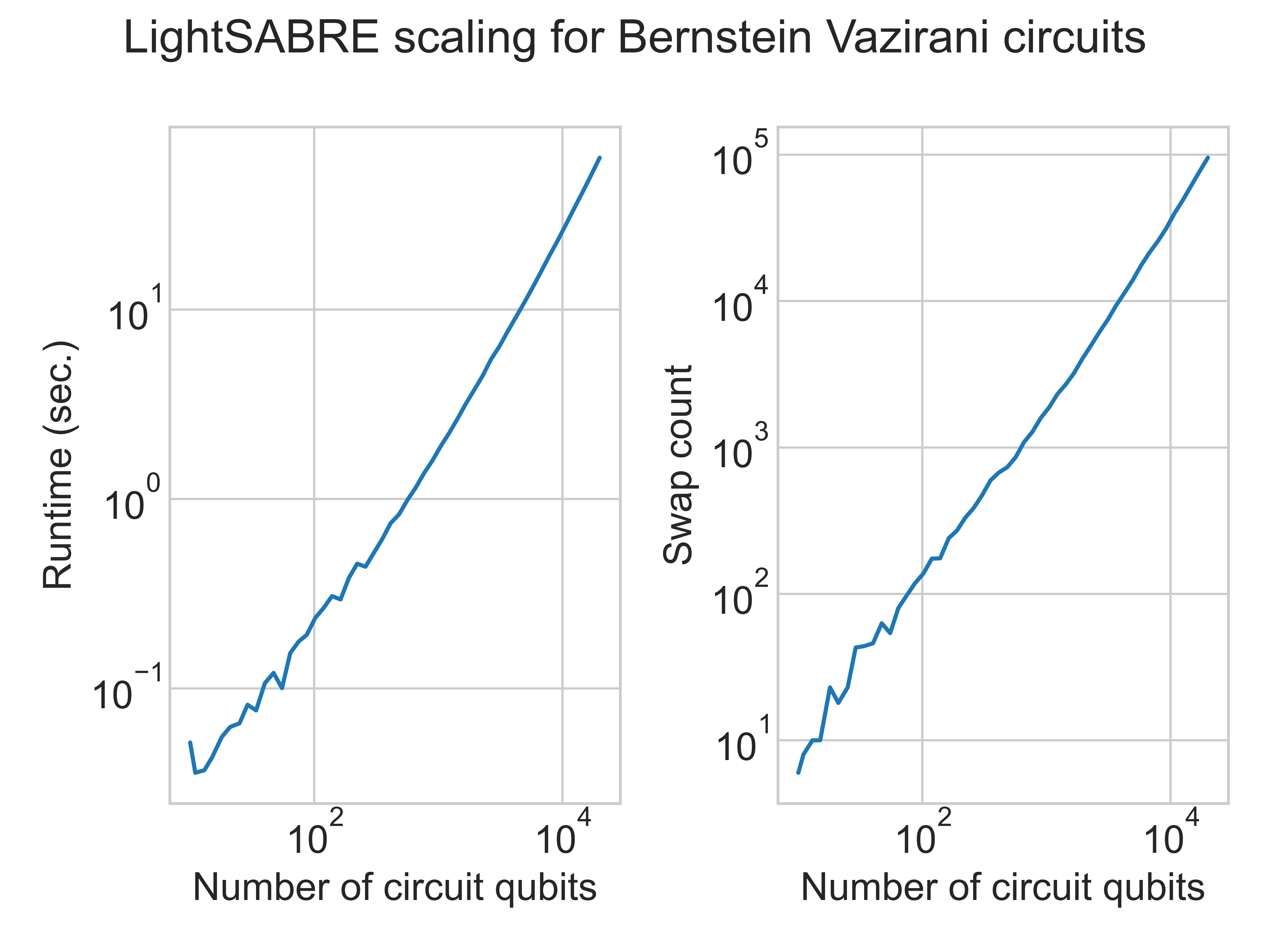}
    \caption{\label{fig:bv_scaling}Algorithm runtime (ignoring setup and output circuit construction 
    as for small circuits these times dominate for such a large connectivity graph) and output swap count for running LightSABRE on 
    Bernstein Vazirani circuits targetting a backend with a 142x142 
    directed grid connectivity with 20 layout and routing trials and 4 iterations. Generated using Qiskit 1.0.2 as bugs introduced in 1.1.0 prevented
    scaling this large. Run using Python 3.12.5 on an AMD Ryzen Threadripper 3970x running Linux 6.10.3.
    }
\end{figure}

When evaluating the scaling performance of routing and layout algorithms, 
it is essential to consider two primary factors: the scaling with respect 
to the number of qubits and the scaling as a function of circuit depth. 
To examine scaling as a function of the number of qubits, Bernstein-Vazirani 
circuits are an ideal benchmark since these circuits scale linearly in gate 
count with the number of qubits.
Figure \ref{fig:bv_scaling} illustrates the runtime performance and the output 
swap gate count from running LightSABRE with 20 layout and 
routing trials on Bernstein Vazirani circuits from 10 qubits to
19998 qubits targetting a backend with a 142x142 directed grid connectivity.
For near-term quantum systems, the runtime scaling of the 
LightSABRE algorithm makes it well-suited for systems with thousands of qubits.

\subsection{Evolution of LightSABRE over time}

\begin{figure}[h!]
  \centering
  
  \begin{subfigure}{\columnwidth}
      \includegraphics[width=\columnwidth]{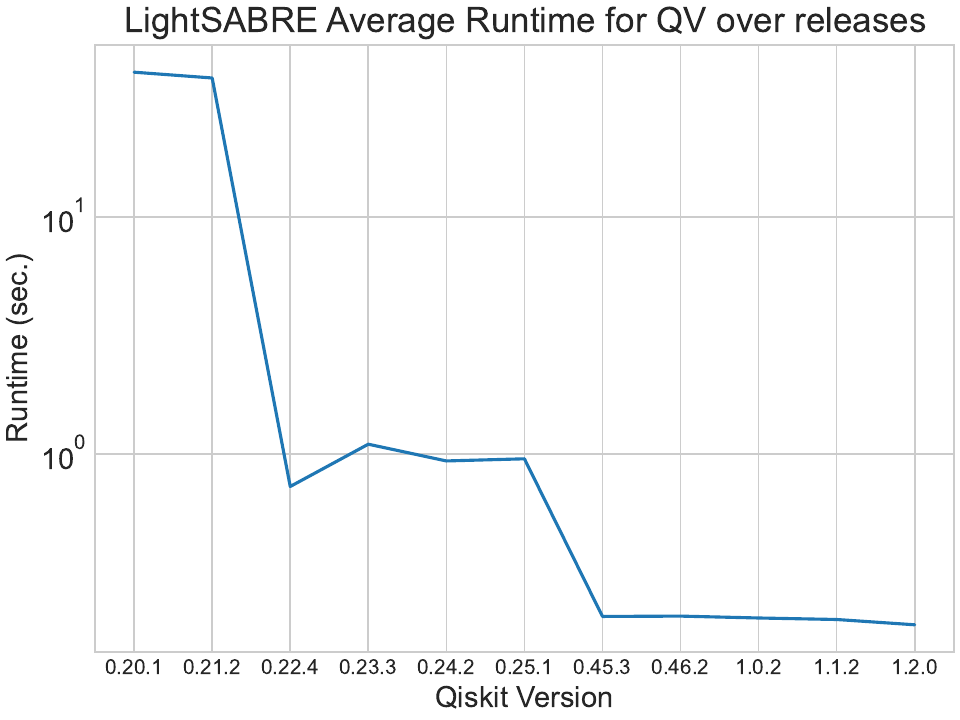}
      \captionsetup{labelformat=empty}
      \caption[FIG. 8a]{FIG. 8a: 
      First parts of SABRE ported to Rust was at 0.22.4, resulting in a significant runtime improvement. 
      From 0.20.1 to 1.2.0, SABRE became approximately 200 times faster.}
      \label{fig:sabre_runtime_over_time}
  \end{subfigure}

  \begin{subfigure}{\columnwidth}
      \includegraphics[width=\columnwidth]{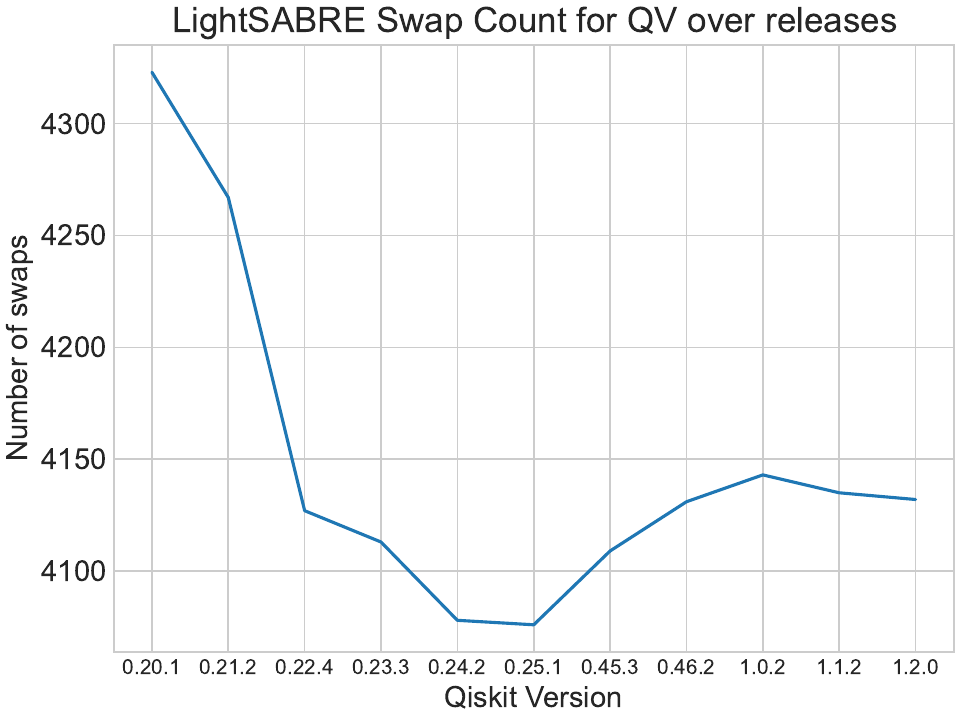}
      \captionsetup{labelformat=empty}
      \caption[FIG. 8b]{FIG. 8b: 
      The randomization of the algorithm likely accounts for the small regression in SWAP 
      count after 0.45.3, as small differences in layout can have a pronounced impact. }
      \label{fig:sabre_swap_count_over_time}
  \end{subfigure}
  
  \caption{\label{fig:sabre_over_time} LightSABRE was significantly optimized, 
  particularly after porting to Rust at Qiskit 0.22.4. 
  All data was generated using 4 iterations, and for Qiskit-terra versions >= 0.23, 
  20 layout and 20 routing trials were run, 
  targeting a 50q QV circuit with 57 qubit heavy-hex connectivity. 
  The tests were conducted on Python 3.9.9 using an AMD Ryzen Threadripper 3970x running Linux 6.10.3.}
\end{figure}

The evolution of the LightSABRE algorithm over time highlights the impact of various techniques introduced in Qiskit, 
as shown in \ref{fig:sabre_over_time}

Figure \ref{fig:sabre_over_time} tracks LightSABRE's performance improvements across Qiskit releases, 
demonstrating the impact of key algorithm refinements. 
These graphs were generated running
Qiskit's LightSABRE pass over the same 50 qubit Quantum Volume \cite{PhysRevA.100.032328} circuit 100 times
with 20 layout and routing trials each and a random seed. 
The graphs starts with qiskit-terra (the legacy package name for what is now Qiskit after 1.0.0) 
in 0.20.1, which introduced the release valve mechanism described in \ref{subsec:release-valve} 
and was the first version capable of completing the example. 
Other releases of highlight are 0.23 introduced multiple trials described in 
\ref{subseq:multiple-trials} and also the relative scoring described in \ref{subsec:relative-scoring}.
The improved in runtime between 0.21.2 and 0.22.4 on the plot is because 0.22.0 was the first 
release where part of the the implementation moved from Python to Rust.

\section{Conclusion}\label{sec:conclusion}

In this work, we have introduced LightSABRE, a significantly enhanced version of the original SABRE algorithm, 
tailored to meet the advancing demands of modern quantum computing. The key result of our enhancements is the 
substantial improvement in runtime, driven primarily by the transition to Rust, which has allowed us to optimize 
performance at a fundamental level. This efficiency gain is critical as quantum devices continue to scale, enabling 
the execution of multiple trials and the exploration of a broader range of potential solutions within a shorter timeframe.

While LightSABRE's primary focus is on improving runtime, it also consistently delivers higher-quality circuits 
through several algorithmic innovations. The introduction of the Relative Scoring mechanism and 
the ability to run multiple trials ensures that LightSABRE not only optimizes for speed 
but also maintains or improves circuit quality, with significant reductions in swap count and depth.

LightSABRE's versatility extends to a wide range of circuit types and optimization goals. 
New heuristic components, such as depth and critical path enhancements, allow for 
fine-tuning of the routing process based on specific performance metrics. LightSABRE’s support 
for disjoint connectivity graphs and classical control flow makes it adaptable to diverse quantum architectures and circuit configurations.

Altogether, LightSABRE represents a major advancement in qubit mapping algorithms, balancing the need 
for speed with the flexibility to achieve high-quality outcomes across a diverse range of quantum circuits. 
As quantum hardware continues to evolve, the improvements embodied in LightSABRE position it 
as a robust and scalable solution for quantum circuit optimization, both in the near term and for future advancements.

\begin{acknowledgments}
We thank Ali Javadi-Abhari, Lev Bishop, and Paul Nation for helpful conversations. This material is based upon work supported by the U.S. 
Department of Energy, Office of Science, National Quantum Information Science Research Centers, 
Co-design Center for Quantum Advantage (C2QA) under contract number DE-SC0012704.
\end{acknowledgments}

\bibliography{refs}
\end{document}